\documentstyle[11pt,newpasp,twoside]{article}
\def\edcomment#1{\iffalse\marginpar{\raggedright\sl#1\/}\else\relax\fi}
\reversemarginpar

\begin{document}
\title{Microlensing in M31 - The MEGA Survey's Prospects and Initial Results}

\author{Arlin Crotts, Robert Uglesich}
\affil{Columbia University, Department of Astronomy, 550 W.~120th St., New
York, NY~~10027, U.S.A.}

\author{Andrew Gould}
\affil{Ohio State University, Department of Astronomy, Columbus, OH 43210,
U.S.A.}

\author{Geza Gyuk}
\affil{University of California, San Diego, Department of Physics, 9500 Gilman
Dr., La Jolla, CA~~92093, U.S.A.}

\author{Penny Sackett, Konrad Kuijken}
\affil{Kapteyn Astronomical Institute, 9700 AV Groningen, Netherlands}

\author{Will Sutherland}
\affil{University of Oxford, Department of Physics, Oxford OX1 3RH, England,
UK}

\author{Lawrence Widrow}
\affil{Queen's University, Department of Physics, Kingston, ON~ K7L 3N6, Canada}

\begin{abstract}
January 2000 completes the first season of intensive, wide-field observations
of microlensing and stellar variability in M31 by MEGA (``{\bf M}icrolensing
{\bf E}xploration of the {\bf G}alaxy and {\bf A}ndromeda'') at the Isaac
Newton 2.5m Telescope, the KPNO 4m, and the 1.3m and 2.5m telescopes of MDM
Observatory.
In preliminary analysis, we detect $\sim$50000 variable objects, including some
consistent with microlensing events.
We present the level of sensitivity to be reached in our planned three-year
program to test for the presence of a significant halo microlensing
population in M31, as well as its spatial distribution and mass-function.
We also discuss our application of
image subtraction to these wide fields and HST WFPC2 Snapshot followup
observations to confirm candidates identified from previous years' surveys.

We present intermediate results from our smaller-field survey, on the MDM
1.3m and Vatican Advanced Technology 1.8m Telescope, from 1994-1998, wherein
we have discovered 8 additional probable microlensing events, over about
one-half the time base of the project, in addition to confirming three of our
original 6 microlensing candidates from 1995.
\end{abstract}

\section{Introduction}

Microlensing has become an interesting probe of dark matter in our Galaxy.
Recent microlensing surveys have indicated that a large fraction of the
matter in the Galactic halo may be dark objects with masses comparable to those
of stars, but have not revealed what these objects might be.
We have detailed how microlensing internal to M31 (Crotts 1992, Gyuk and Crotts
2000) might be used
to test such results and show better how the microlensing matter is
distributed in space and as a function of mass.
A survey of small fields in M31 has revealed several such candidate events, at
roughly the predicted rate (Crotts \& Tomaney 1996).
We discuss below what efforts have been required to further verify the
microlensing nature of these events.

Many papers predict microlensing optical depths $\tau$ in M31, which
should approach $\tau \approx 10^{-5}$, over ten times greater than
towards the LMC, but previously none of these works studied variations in halo
microlensing optical depth over the face of M31.
Since M31 differs from the Galaxy in that many sightlines for microlensing
are seen by an observer at Earth, the variation of $\tau$ depending on
the spatial distribution of microlensing objects should be explored.

\section{Testing Candidate Microlensing Events}

Originally, Crotts \& Tomaney (1996) identified six events from the 1995
observing season in M31 (using the Vatican Advanced Technology 1.8-meter
telescope on Mt.~Graham, Arizona) over $\sim60$d covering a 125 arcmin$^2$
field.
These events were characterized by full-width half-maximum timescales of
10d$<t_{fwhm}<50$d.
The longer end of this range is troublesome, coinciding with
pulsewidths seen in miras and other longterm red variables.
In fact, a mira lightcurve, resembling a symmetric sawtooth in magnitude, with
peak-to-valley amplitudes of about 5 magnitudes in the $R$ band, appears
similar to the lightcurve of a simple (point-mass, point-source) microlensing
event during its maximum amplification.
Furthermore, if the period of a mira is about $2 \over 3$~yr, one must monitor
for two M31 seasons beyond (or preceding) the peak in the light curve in order
to detect another peak, since proximate peaks occur when M31 is not easily
observable.
The peak of such a mira has $t_{fwhm}\approx 40$d, which would corresponds to
lensing masses $m \approx 0.5 M_\odot - 1 M_\odot$ for typical lensing
geometries in our survey region.
Hence mira-like variables are troublesome contributors to a potential false
event rate, and require multiple seasons of observations in order to be
eliminated.

We have performed two tests of these six original candidates: 1) constructing
well-sampled lightcurves over the M31 seasons of the three subsequent years
(through 1998),
and 2) obtaining $HST$ WFPC2 snapshot observations of these sources in order to
determine if their colors are consistent with mira-like variables.
(The latter test is impossible from the ground, since crowding does not allow
one to resolve typical sources in average seeing conditions.
Variable sources are made to appear isolated from one another by virtue of
image subtraction e.g.~Tomaney \& Crotts [1996]).
The result of these two event filters is to eliminate three of the six
events, with the remainder firmly inconsistent with mira-like variables.
For the remaining events, now that we have measured their baseline magnitudes
from WFPC2 images, we can calculate a more accurate peak amplification
(assuming that the lensing mass rests as close as possible along the sightline
to the core of M31).
These persist with estimates for the lensing masses in the range $0.3M_\odot
\la m \la 2 M_\odot$, with two events possible arising from stars in M31's
bulge, but one almost certainly not a bulge lens, given its source position
2.5~kpc out into the disk.

\section{Possible Results from a Larger Survey}

Given the robust nature of some of the candidate microlensing events from the
small area survey discussed above, it is worth considering possible results of
a larger, wide-angle survey, especially since the advent of CCD imagers
covering large fractions of a square degree.
We present here representative results simulating a survey in which roughly
0.5 square degree is imaged for two hours every night on a
two-meter telescope, in 1-arcsec seeing, requiring each event to be sampled at
the 4$\sigma$ level over at least 3 day timescales.
The event rate predicted for such a survey is large e.g.~Figure 1, which shows
the predicted distribution of events over the field containing our survey area,
for a typical model with 50\% of the halo dark matter composed of 0.5~M$_\odot$
microlensing masses.

\begin{figure}
\plotfiddle{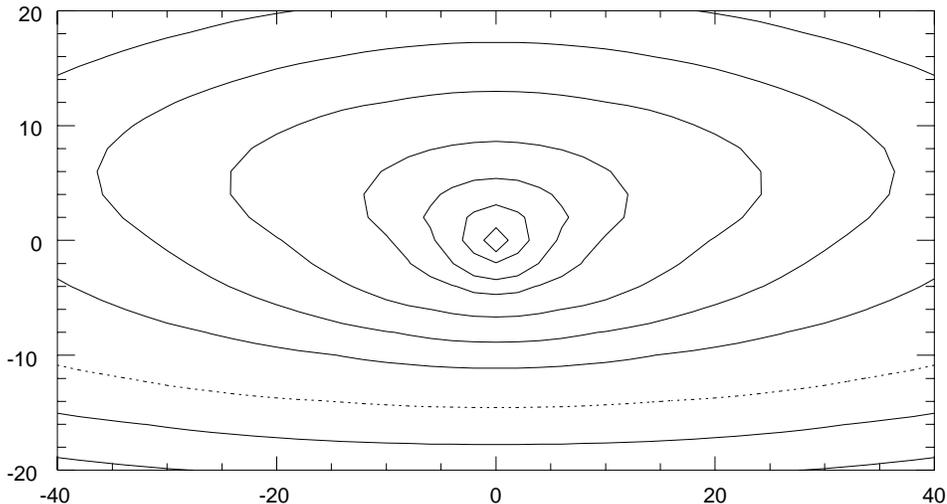}{2.35in}{-90}{060}{060}{-265}{325}
\caption{
An 80$\times$40 arcmin$^2$ plot of M31's center (major axis of M31 is
horizontal, minor axis vertical) showing contours of the predicted event rate
for bulge and halo microlensing in M31.
The highest contour is for 50 events yr$^{-1}$ arcmin$^{-1}$ (near the center),
with lower contours at 20, 10, 5, 2, 1, 0.5, 0.2 (dotted), 0.1 and 0.05 events
y$^{-1}$ arcmin$^{-1}$.
This reasonable model, for an unflattened halo with a 5~kpc core radius,
predicts over 100 detections during an M31 observing season.
}
\end{figure}

The halo fraction, halo flattening ($q$) and core radius ($r_c$) are allowed to
vary between models, and then a maximum likelihood calculation is performed to
yield resulting values for these parameters.
These parameters can result in large changes in the distribution of
microlensing optical depth across the face of M31 (see Figure 2), which
significantly affects the distribution of microlensing event detections.

\begin{figure}
\plotfiddle{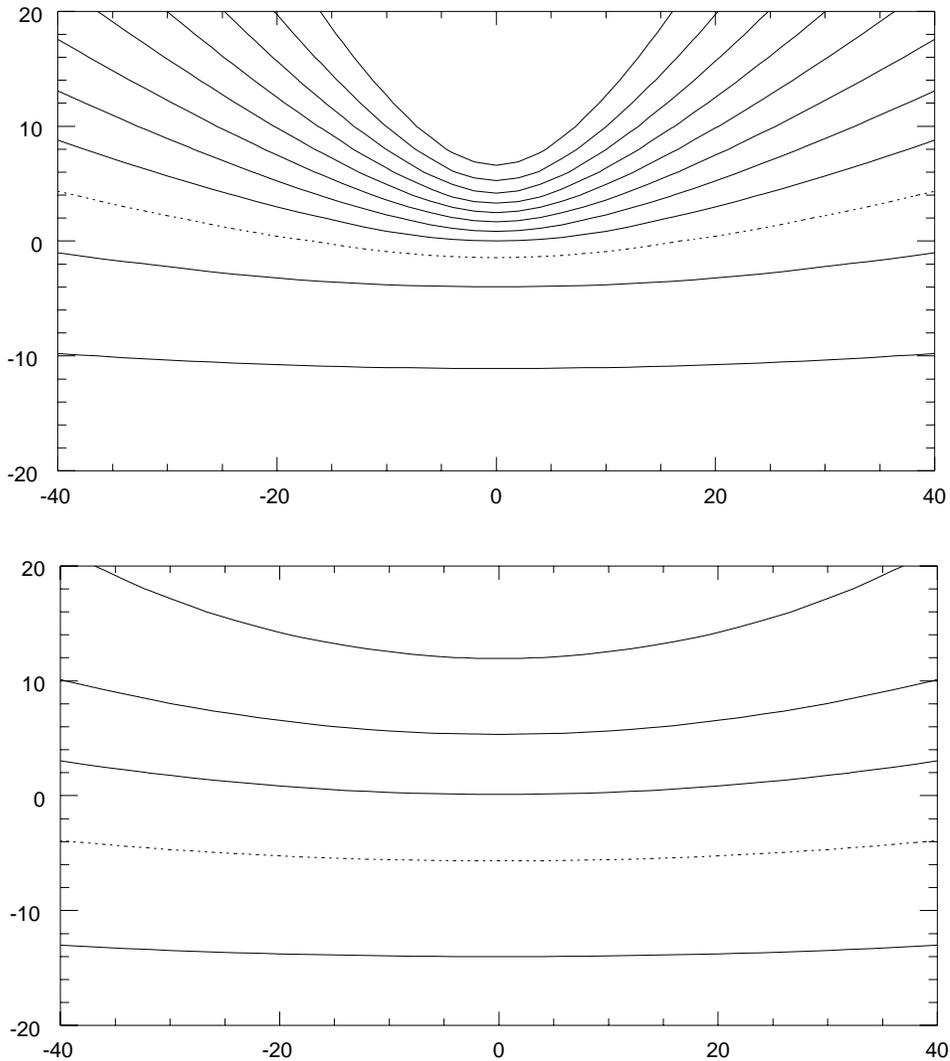}{5.5in}{-90}{060}{060}{-266}{540}
\caption{
The same region as in Fig.~1, but for contours of lensing optical
depth (halo only) for two models of the spatial distribution of halo
microlensing masses.
The top panel corresponds to an unflattened model with 1.5~kpc core
radius, and, on the bottom, a 3.3-to-1 flattening and a 10~kpc core radius.
The dotted contours (just below center in each panel) correspond to
$2\times 10^{-6}$, increasing towards the top (far side of
disk) to over $6\times 10^{-6}$ in the top panel and $3.5\times 10^{-6}$ in
the bottom.
In addition to the halo contribution shown, the central 10 arcmin diameter
contains
a $\tau \la 5\times 10^{-6}$ bulge signal (represented in Fig.~1); a uniform
$\tau \la 10^{-6}$ due to the Galaxy and a much smaller contribution from
M31 disk-disk lensing (Gould 1994) are spread throughout.
}
\end{figure}

\begin{figure}
\plotfiddle{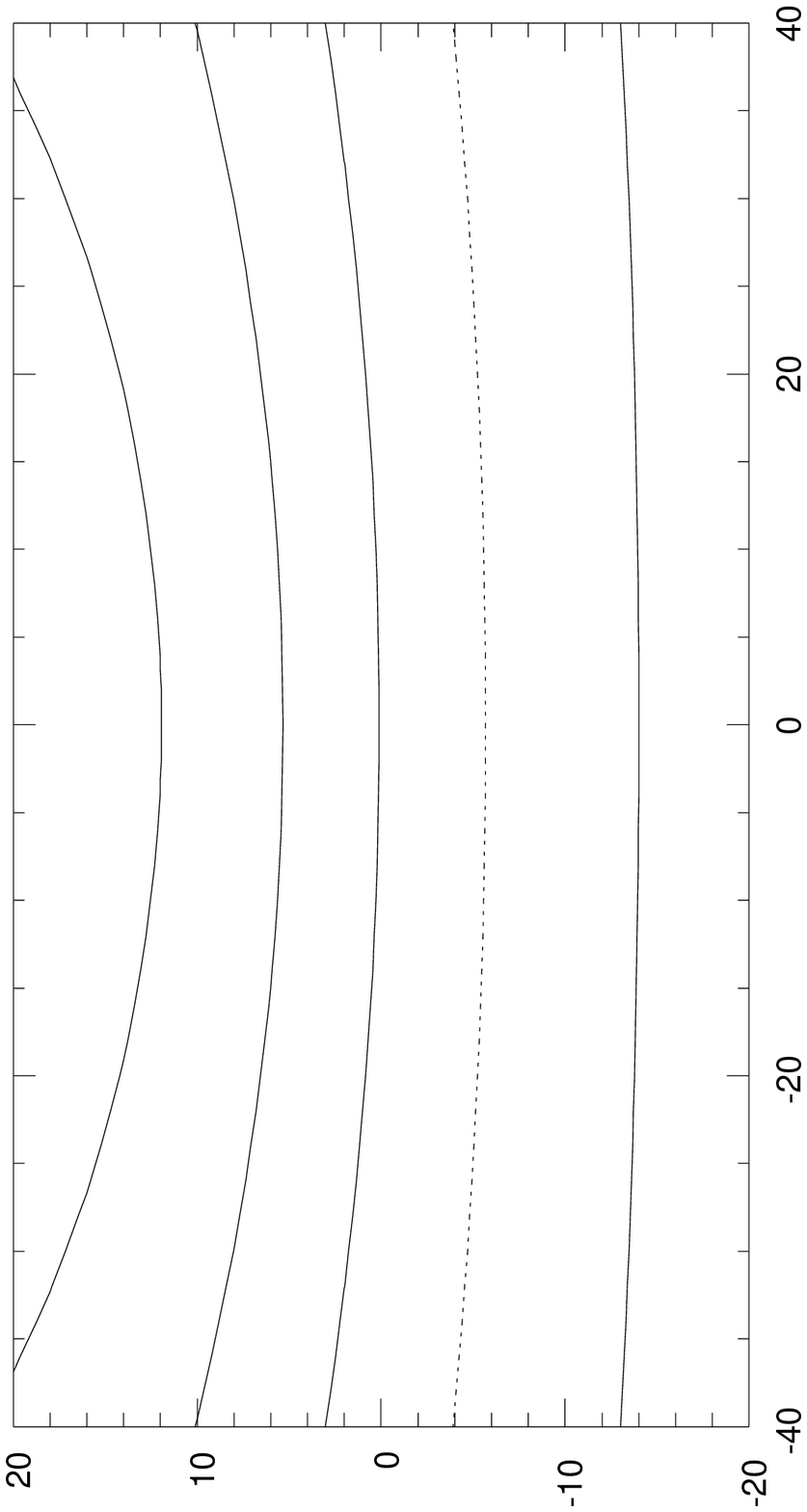}{0.0in}{-90}{060}{060}{-265}{510}
\end{figure}

\pagebreak

Our calculations show that this larger survey might easily observe $\sim$100
such events per M31 observing season, which would allow the shape of a strong
microlensing halo of M31 to be mapped.
Since most masses reside near where the sightline passes the center of the
galaxy, at a known source-lens distance, this survey would also allow a more
exact determination of the masses doing the lensing.
Selected fields in M31 might also serve as independent sightlines through the
halo of our Galaxy.
The preliminary epochs for a large survey in M31, over one-half square degree,
have already been obtained, initiating the project MEGA: Microlensing
Exploration of the Galaxy and Andromeda.

\begin{figure}
\plotfiddle{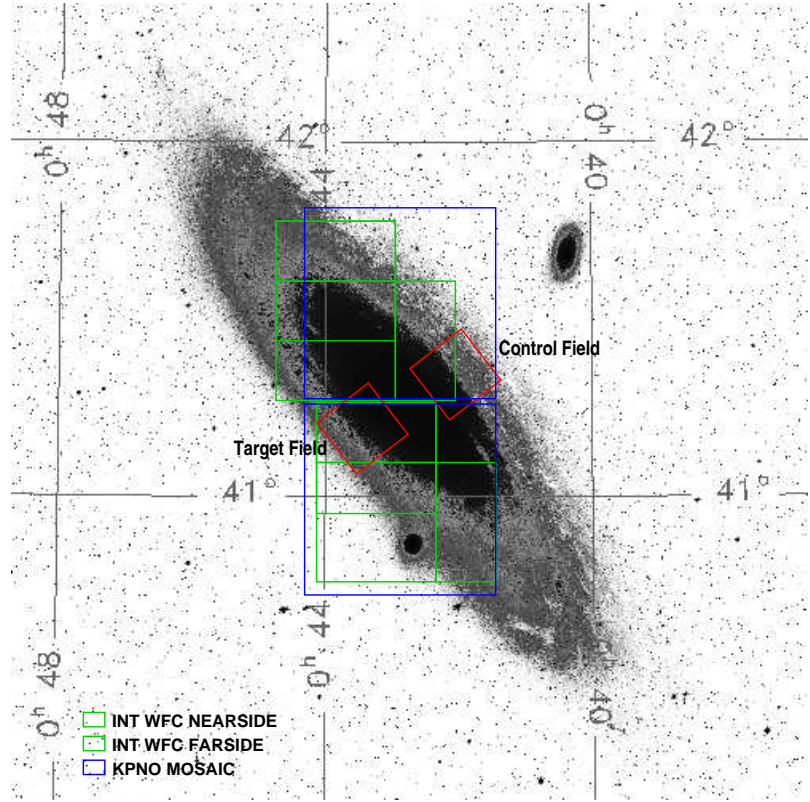}{4.0in}{0}{060}{060}{-180}{-090}
\caption{
MEGA fields being monitored on the INT WFC (complex, light-shaded polygons)
and KPNO 4m/MOSAIC (large, dark-shaded squares), as well as the VATT/Columbia
fields (smaller, diagonal squares: ``Target'' and ``Control'').
MDM fields are not shown.
}
\end{figure}

Our ability to measure $r_c$ and $q$ depend on the true value of $r_c$, with
small values providing greater $\tau$ in the galaxy's center, where more
sources exist.
All models produce $\ga$100 events per season, with small $r_c$ models
producing more.
After three seasons, $r_c$ can be measured to within $\sim$1.5~kpc (1$\sigma$),
and $q$ to $\sim$0.1 (for $r_c < 5$~kpc), or $\sim$2.5~kpc and $\sim$0.2,
respectively for $r_c > 10$~kpc.
With $q$ and $r_c$ well-constrained, the data allow a superior estimate of lens
mass distribution.
This many events can result from a campaign using existing wide-field CCD
arrays on two-meter$+$ class telescopes.
Figure 3 shows the fields being covered by MEGA on some of the telescopes
being used, compared to the fields for the earlier VATT/Columbia survey
(described below).
We have initiated this effort (MEGA) by establishing long baselines
eliminating long-period variables, having
obtained several epochs of such data in 1997 and 1998, and having begun more
intensive observations in 1999 to detect microlensing events across much of
M31 over the course of several seasons.
At the time of the meeting, we have found approximately 40000 variables in the
INT WFC data (some consistent with microlensing), which implies a sample of
some 50000 when KPNO and MDM fields are included.
(INT data are being collected in cooperation with AGAPE - see Kerins, this
volume.)

Considerable improvements can be made in estimating halo shape parameters, as
well as the microlens mass, if Einstein crossing times $t_{ein}$ can be
measured for events.
Since the densest portion of the halo should sit over the bulge of M31, the
lens-source distance is constrained by noting the impact parameter of the line
of sight to the source relative to M31's center.
Hence, $t_{ein}$ can be converted, roughly, to a mass, at a much greater
accuracy than in the LMC sightline situation.
Furthermore, events that are due to halo lenses have preferentially longer
$t_{ein}$ than confusing bulge events, hence measurements of $t_{ein}$ can lead
to better determination of halo shape parameters beyond those accuracies quoted
above.

The Einstein crossing time can be measured in two ways, the $t_{\sigma n}$
method of Baltz and Silk (2000), and by using $HST$ imaging to determine
baseline magnitudes by resolving the source star.
As we showed at the meeting, neither of these is sufficiently reliable by
itself to determine $t_{ein}$ without a large fraction of outlying
measurements, either due to misidentification of source stars in $HST$ images,
or poor $S/N$ in the wings of microlensing lightcurves, where $t_{\sigma n}$ is
determined.
However, if a large field ($\sim 0.2$ deg$^2$) is imaged with $HST$/ACS,
requiring $\sim$ 15 orbits, both methods will be available ($t_{\sigma n}$ from
ground-based data), leading to $t_{ein}$ for $\sim 100$ events, constraining
the average microlensing mass to $\sim 0.1$M$_\odot$ or better, and
significantly improving halo shape parameter determinations.

\section{Intermediate Results from the VATT/Columbia Survey}

Uglesich, Crotts and Tomaney conducted a preliminary survey of two 125
arcmin$^2$ fields using the VATT 1.8m and MDM 1.3m.
Data are reduced using the difference image photometry method of Tomaney and
Crotts (1996), and example of which is shown in Figure 4 for typical MDM 1.3m
images.
The $S/N$ ratio in these difference images is limited by photon shot noise, not
seeing fluctuations, as evidenced in several studies e.g.~Uglesich et al.~1999.
Quantities of data sufficient to find microlensing events for masses in the
0.1-1M$_\odot$ range where obtained in late 1995 through early 1999, with the
best temporal coverage in the 1998-1999 season.
Most of these data are now reduced, with the final season now having been
searched for possible microlensing events.
There are still $\sim$20 epochs to be added from additional observatories, but
already 11 potential events have been found, with 8 sampled over the peak of
the best-fit microlensing curve, and points on either side.
These 8 candidate events are shown in Figure 5.
(At the time of the meeting, we had only fit these with the high-amplification
microlensing curve fit of Gould [1997], rather than the full parameter range
implicit in general point-mass, point-lens microlensing curves of Paczynski
[1986].)~
These sources have also been monitored in previous years' data, and have
maintained a stable baseline, inconsistent with longterm variables e.g.~miras.
We expect a large number of events to be found in this season's and other
season's data.
The VATT/Columbia survey has the potential for sensing the asymmetry in
microlensing events across the face of M31, as seen in Figure 2, if a halo
microlensing population actually exists as a significant fraction of the dark
matter in spiral galaxy halos.

\acknowledgements

This research is supported by NWO grant GBE-M 612-21-013, STScI grants GO 7376
and AR 7970, and NSF grants AST 95-29273, 97-27520, 98-02984, 00-70882 and INT
96-03405, and NSERC (Canada).

\begin{figure}
\plotfiddle{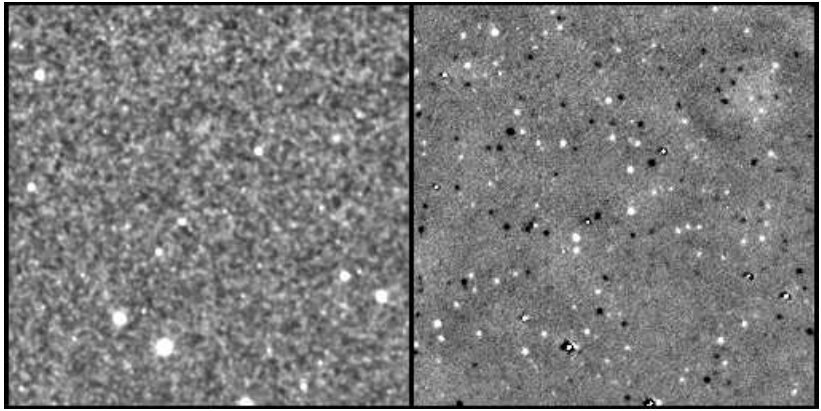}{2.2in}{0}{060}{060}{-190}{-160}
\caption{
MDM 1.3m CCD subimage in MEGA field, processed with
our image subtraction technique.
Left panel shows 110 arcsec section from ``raw'' image taken 28 Oct 1998,
a 3h exposure in $\sim$1~arcsec seeing.
Right panel shows results subtraction of average from the 1997 season of MDM
data.
A few saturated stars and $\sim 250$ variables are seen.
}
\end{figure}

\begin{figure}
\plotfiddle{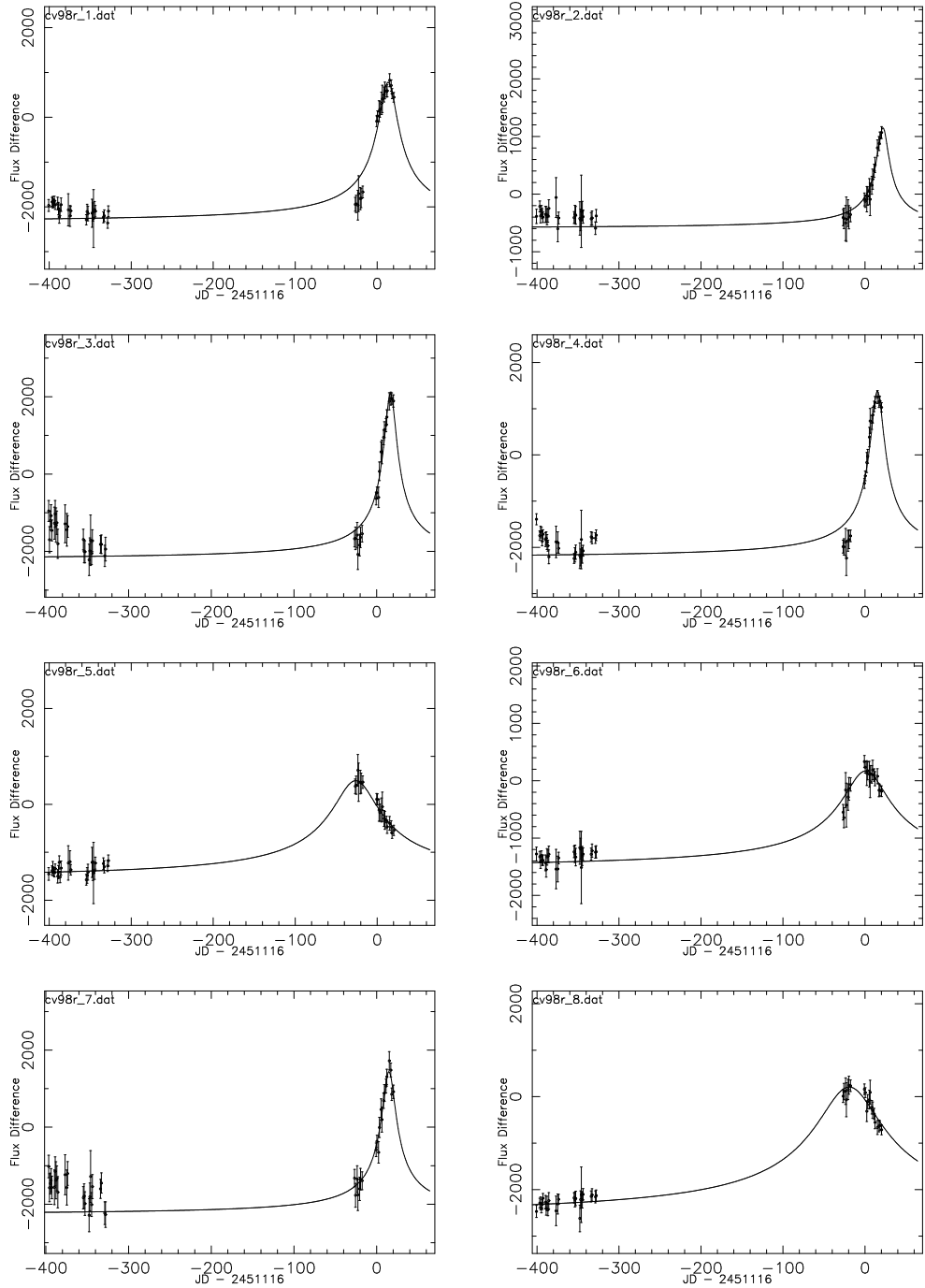}{6.9in}{-00}{070}{070}{-215}{-020}
\caption{
Eight new microlensing candidate lightcurves (ADUs vs.~JD) from 1998
R-band MDM 1.3m data.
Data from other observatories (and MDM) show that these do not vary on
extended baselines, and will fill in points during each event.
Curves fit to the data are from Gould (1997) high-amplification fits, not
full Paczynski (1986) fits.
}
\end{figure}

\end{document}